\documentstyle[eqsecnum,preprint,aps,prd]{revtex}
\def\PB#1#2{\Big\{#1, #2\Big\}}

\def\be{\begin{equation}}
\def\ee{\end{equation}}
\newcommand{\bea}{\begin{eqnarray}}
\newcommand{\eea}{\end{eqnarray}}
\def\we{\approx}
\def\rla{\leftrightarrow}
\def\fr{\frac}
\def\a{\alpha}
\def\b{\beta}
\def\c{\gamma}
\def\d{\delta}
\def\e{\epsilon}
\def\f{\varphi}
\def\f{\phi}
\def\g{\gamma}

\def\l{\lambda}
\def\m{\mu}
\def\n{\nu}

\def\w{\omega}
\def\W{\Omega}

\def\d{\delta}

\def\L{\Lambda}
\def\CL{\cal L}

\def\p{\partial}

\def\T{\Theta}
\def\nn{\noindent}
\def\no{\nonumber}

\begin{document}
\draft
\preprint{\vbox{\hbox{IP/BBSR/2001-6}}}
\title{Born-Infeld Chern-Simons Theory: Hamiltonian Embedding, Duality and
Bosonization}
\author{ E. Harikumar$^1$, Avinash Khare$^2${\thanks{khare@iopb.res.in}}, 
 M. Sivakumar$^1${\thanks{mssp@uohyd.ernet.in}} Prasanta K. Tripathy$^2$,}
\address{$^1$School of Physics, University of Hyderabad,Hyderabad-500
046, INDIA\\ $^2$Institute of Physics, Sachivalaya Marg, Bhubaneswar-751
005, INDIA.} 
\maketitle
\begin{abstract} 
In this paper we study in detail the equivalence of the recently
introduced Born-Infeld self dual model to the Abelian
Born-Infeld-Chern-Simons model in $2+1$ dimensions. We first apply the
improved Batalin, Fradkin and Tyutin scheme, to embed the Born-Infeld Self
dual model to a gauge system and show that the embedded model is
equivalent to Abelian Born-Infeld-Chern-Simons theory. Next, using
Buscher's duality procedure, we demonstrate this equivalence in a
covariant Lagrangian formulation and also derive the mapping between the
n-point correlators of the (dual) field strength in Born-Infeld
Chern-Simons theory and of basic field in Born-Infeld Self dual model.
Using this equivalence, the bosonization of a massive Dirac theory with a
non-polynomial Thirring type current-current coupling, to leading order in
(inverse) fermion mass is also discussed.  We also re-derive it using a
master Lagrangian. Finally, the operator equivalence between the
fermionic current and (dual) field strength of Born-Infeld Chern-Simons
theory is deduced at the level of correlators and using this the
current-current commutators are obtained.
\\

\end{abstract}
\medskip\pacs{PACS Numbers: 11.10.Kk, 11.15.-q, 11.10.Ef, 11.40.-q\\ 
Keywords:\,Born-Infeld, Chern-Simons, Hamiltonian Embedding, Duality,
Bosonization.\\} 
\newpage

\section{Introduction}\label{intro}

The Born-Infeld theory has been an area of intense activity in the recent
times because of the important role it plays in the context of string
theory \cite{Tseytlin}. Some time back, after the discovery of D-branes,
it was realized that the dynamics of the gauge field excitations on world
volume of D-branes is described by the Born-Infeld theory. Various aspects
of the Born-Infeld theory have been studied thoroughly both from the
string as well as field theoretic point of view. One of the most remarkable
properties of the D-branes is that they carry Ramond-Ramond (RR) charges and 
hence, must couple to the RR states of the closed string. These couplings 
are incorporated via the Chern-Simons action, which is constructed 
from the antisymmetric combination of the field strength with the 
pull back of the RR gauge potential on to the world volume of the brane.
The coefficient of the Chern-Simons term gives rise to the RR charge of
the brane. Thus dynamics of $Dp$-brane is described by sum of Born-Infeld 
and Chern-Simons actions in $p+1$ dimensions. In particular, the $D2$-brane
dynamics is described by 3 dimensional Born-Infeld Chern-Simons (BICS) action. 
It may thus be of some interest to examine 
the various aspects of BICS theory in 
$2+1$ dimensions which is what we propose to do in this paper .

Several years ago Deser and Jackiw \cite{Deser:1984pl} showed that the
self-dual model given by
\cite{Townsend:1984pl} 
\be
{\CL}=\fr{1}{2}f_\m f^\m -\fr{1}{2m}\e_{\m\n\l}f^\m\p^\n f^\l~,
\label{sdl}
\ee
is equivalent to the Maxwell-Chern-Simons (MCS) theory given by 
\be
{\CL}=-\fr{1}{4}F_{\m\n}F^{\m\n} +\fr{m}{2}\e_{\m\n\l}A^\m \p^\n A^\l~.
\label{mcs}
\ee
The equivalence between the two models has been extensively studied in the
literature \cite{{Banerjee:1997pr},{Kim:1998ijm}} and has found application in
studying bosonization
in $2+1$ dimensions \cite{{Redlich:1984prd},{Fradkin:1994pl},{Rabin:1995pl}}.

Recently two of us extended this equivalence to the Born-Infeld case
\cite{Prasanta:2001pl}. In particular, a self dual model for the 
Born-Infeld case (SDBI) described by the Lagrangian
\be
{\CL} = \b^2 \sqrt{1+\fr{1}{\b^2} f_\m f^\m} -
\fr{1}{2m}\e_{\m\n\l}f^\m\p^\n f^\l~,
\label{bisd1} 
\ee
was introduced and was shown to be equivalent to the BICS theory\cite{Prasanta:2001pl} 
as given by
\be{\CL} = \b^2 \sqrt{1-\fr{1}{2\b^2} F_{\m\n}F^{\m\n}}
+\fr{m}{2}\e_{\m\n\l}A^\m \p^\n A^\l~.
\label{bimcs1}
\ee

It may be noted that the action corresponding to the BICS Lagrangian is
invariant under the $U(1)$ transformation of the vector field
\be
A_\m\to A_\m+\p_\m \l~,
\label{au1}
\ee
while the SDBI Lagrangian (\ref{bisd1})  as well as the corresponding
action is not gauge
invariant. As expected, the BICS Lagrangian (\ref{bimcs1}) reduces to that
of the Maxwell-Chern-Simons theory(MCS) (\ref{mcs}) while the SDBI
Lagrangian (\ref{bisd1}) goes over to the SD Lagrangian (\ref{sdl}) in the
limit $\b^2 \to \infty$.

This raises the question of extending the previous analysis regarding the
equivalence of the MCS and SD models to the Born-Infeld case and in
particular study the bosonization issue. Further, it is interesting to
enquire the role of non-linearity in the Born-Infeld part of the theory in
extending the earlier studies of MCS theory.

At this point we recall that the previous studies of the equivalence
relied on converting the second-class constraints in the self-dual model
to the first class constraints by applying Batalin, Fradkin and  Tyutin
(BFT) procedure\cite{{Batalin:1987fm},{Batalin:1991jm}}. The inherent
non-linearity in Born-Infeld theories makes the application of this method
interesting and somewhat nontrivial. There also exists a method due to
Buscher \cite{{Buscher:1987pl},{Rocek:1992np}} for obtaining duality
equivalence between different theories, which has been applied to the MCS
theory and the corresponding self-dual model\cite{us} has been obtained. 
It should
again be interesting to apply this scheme to the BICS theory.

The role of Chern-Simons gauge field in 2+1 dimensional bosonization is 
well studied for the case of matter field coupled with Chern-Simons gauge field
\cite{{Shaji:1990ij},{Shankar:1991ij}}
and also for a generic current-current interacting
theory\cite{Barci:1999prd}. 
Specifically using the connection of the MCS gauge theory to the self-dual
model, the MCS gauge theory has been shown to be equivalent to the massive
Thirring model up to leading order in (inverse) fermionic
mass\cite{Fradkin:1994pl}. It is then natural to enquire if a similar
study is also possible in the case of the BICS gauge theory.

In this paper we study three aspects of the BICS-BISD correspondence,
which are:
\begin{enumerate}
\item Hamiltonian embedding of BISD theory and its equivalence to the BICS
theory 
\item the equivalence as a duality relation using Buscher's procedure
\item using it to study the bosonization of a massive interacting
Dirac theory with a non-polynomial current-current interaction.
\end{enumerate}

We start with the BISD model which has only second-class constraints  and
convert them to first class constraints and also the Hamiltonian to a
gauge invariant one following the  generalized BFT scheme. Then we show
that the embedded model is equivalent to the BICS theory.

We also show the equivalence between the BICS theory and the BISD model in
a manifestly covariant manner, by applying Buscher's procedure, which
relies on the presence of a global symmetry in the theory. Here we show
that these two theories are related to each other by a duality mapping. We
also provide a mapping between the fields at the level of correlators
between the two theories.

We next use the equivalence between the BISD and the BICS to show that 
to the leading order in the inverse fermion mass $\mu$, the
latter is a bosonized version of a massive non-polynomial Thirring (BIT)
type current-current interacting theory described by the Lagrangian
\be
{\CL}= {\bar \Psi}(i\c_\m \partial^\m-\m)\Psi + \b^2 \sqrt{
1-\fr{\l}{\b^2} j_\m j^\m}~.
\label{bith1}
\ee
Here $j_\m$ is the fermionic current ${\bar\Psi}\c_\m\Psi$ and 
$\l$  and $\beta$ are dimensionful constants. This is also re-derived
starting from a master Lagrangian and operator correspondence between the
fermionic current of the BIT theory and (dual) field strength of the BICS
theory at the level of correlators is also obtained.
We also evaluate the current algebra of the
fermionic theory using the operator correspondence. It is found that even
though the current algebra is $\b$ dependent (of course in the limit
$\b^2\to\infty$ it agrees with massive Thirring model result) but the
bosonization rules are $\b$-independent, hence they are the same for the
MCS-Thirring \cite{Rabin:1995pl} and the BICS-BIT cases.

This paper is organized in the following way. In the next section, we
apply the Hamiltonian embedding to the BISD model and show that the
embedded model is equivalent to the BICS theory. In section III, we apply
Buscher's procedure to show the equivalence of the BICS theory and the
BISD model in the Lagrangian formulation. Here we also derive the mapping
between the n-point correlators of these two models. In section IV, we
derive the mapping between the fermionic BIT theory and the bosonic 
BICS theory and
obtain the bosonization rules as well as the current algebra of the
fermionic current. We conclude with discussions in section V. In the
Appendix \ref{a1}, using the symplectic quantization scheme, the Dirac
brackets of the BICS gauge theory are obtained which are needed for
the current algebra evaluation.

We work with $g_{\m\n}=diag (1,-1,-1)$ and $\e_{012}~=~\e^{012}~=~1.$

\section{Hamiltonian Embedding}\label {embed}

In this section, we apply the improved Hamiltonian embedding procedure of
Batalin, Fradkin and Tyutin (BFT)\cite{Yong:1997ij} to the BISD model and
convert it into a gauge theory. Then, using the solution of the embedded
Hamilton's equation we show the equivalence of the embedded model to the
BICS theory. First, we briefly sketch the Improved BFT procedure and then
we apply it to the BISD model.

\subsection{Improved BFT Embedding}

In the Batalin, Fradkin and Tyutin (BFT) method, first one enlarges the
phase space by introducing auxiliary variables $\Phi_\a$ corresponding to
each of the second class constraints $(T_{\a}(P, Q))$ (where $P$ and $Q$
stand for the original canonically conjugate phase space variables)
satisfying, 
\be
\{\Phi_\a,\Phi_\b\} =\w_{\a\b}~,
\label{bft.pb}
\ee
such that ${\rm det}|\w_{\a\b}|\ne0$ and $\w_{\a\b}$ is field
independent.  Here the new variables $\Phi_\a$ and the original second
class constraints $(T_{\a}(P, Q))$ are of the same Grassman parity.  Now
we define the first class constraints ${\bar T}_{\a}(P, Q, \Phi_\a)$ in
the extended phase space, satisfying 
\be
\{{\bar T}_\a, {\bar T}_\b\}=0~,
\label{fpb}
\ee
and the solution for which is obtained as 
\be
{\bar T}_\a=\sum_{n=0}^{\infty} T_{\a}(n)~,
\label{sol.pb}
\ee
 where $n$ is the order of the term in $\Phi_\a$ and $T_{\a}(0)=T_\a$.

After converting the second class constraints to strongly involutive ones,
one proceeds to construct a gauge invariant Hamiltonian ${\bar
H}(P,Q,\Phi_\a)$ in the extended phase space. This gauge invariant
Hamiltonian has to satisfy 
\be
\{{\bar T}_\a, {\bar H}\}=0~,
\label{bft.cod1}
\ee
whose solution is obtained as an infinite series,
\be
{\bar H}=\sum_{n=0}^{\infty} H_n~,~~~{\rm with}~~H_0=~ H~,
\label{hsol}
\ee
where $H={\bar H}(P,Q, \Phi_\a=0)$.  In the case of the linear theories, it is
found that the series (\ref{sol.pb}) and (\ref{hsol}) have only a finite
number of terms (i.e., $n$ is finite). However, in the case of the non-linear
theories, this series need not terminate. Also, one may not be able to
express the series in a closed form and this makes the implementation of
the procedure rather complicated. This will be shown to occur to the BISD
model in section IIB. In the improved BFT formalism one can circumvent
this problem as follows. Corresponding to each of the original phase space
variables $\f$, one constructs ${\bar \f}$ in the enlarged phase space,
satisfying
\be
\PB{{\bar T}_\a}{{\bar \f}}=0~.
\label{imp.cond}
\ee
Thus ${\bar \f}$ is a gauge invariant combination.  Now by replacing $\f$
with ${\bar \f}$ in any function $A(\f)$ of the phase space variable, one
can obtain the corresponding involutive function ${\bar A}$ in the
enlarged phase space.

We now apply the improved BFT formulation to the BISD model.
First we linearize the BISD model by introducing a multiplier field. Then
in the linearized form, apart from the original constraints we have
constraints coming from the linearizing field. Since we plan to eliminate
this field after carrying out the embedding, we concentrate only in the
sector of original constraints. We apply a partial
embedding of the BISD model using the improved BFT scheme. Using the
solutions of the embedded Hamilton's equations, we map the embedded model
to the BICS theory.

\subsection{Improved BFT Embedding of Born-Infeld Self Dual Model}

We start from the Lagrangian
\be
{\CL}=\Phi +\fr{\b^4}{4\Phi} + \fr{\Phi}{\b^2} f_\m f^\m
-\fr{1}{2m}\e_{\m\n\l}f^\m\p^\n f^\l~.
\label{bisd.lag}
\ee
Note that by eliminating the auxiliary $\Phi$ field from (\ref{bisd.lag}) 
by using 
its equation of motion gives back the standard BISD Lagrangian of Eqn.
(\ref{bisd1}).

\nn The primary constraints following from the above Lagrangian
(\ref{bisd.lag}) are
\bea
\Pi_0~\we~0,~~~\Pi_{\Phi}~\we~0,\label{prim}\\
\W_i~=~(\Pi_i +\fr{1}{2m}\e_{0ij}f^j)~\we~0~.
\label{symp}
\eea
Here $~\Pi_{\m} {\rm ~and~} \Pi_{\Phi}~$ are the conjugate momenta
corresponding to $~f_\m {\rm ~and~} \Phi$ respectively.  The canonical
Hamiltonian is
\be
H_c = -\fr{\Phi}{\b^2} f_i f^i -\Phi -\fr{\b^4}{4\Phi} -f_0
(\fr{\Phi}{\b^2}f^0 -\fr{1}{m}\e_{0ij}\p^i f^j)~.
\label{bisd.ham}
\ee 
The persistence of the primary constraints leads to the following
secondary constraints
\bea
\W= \fr{2 \Phi}{\beta^2} f_0 -\fr{1}{m}\e_{0ij}\p^i f^j \we 0~,\label{sec1}\\
\L= 1 - \fr{\b^4}{4{\Phi}^2} +\fr{1}{\b^2} f_if^i +\fr{1}{\beta^2} f_0f^0\we
0~.
\label{sec2}
\eea

In the above, the constraint $\W_i$~(\ref{symp}) is due to the symplectic
structure of the theory and not a true constraint. Following Faddeev and
Jackiw\cite{Faddeev:1988pr}, we impose this symplectic condition strongly
leading to the modified bracket
\be
\PB{f_i(x)}{f_j(y)}=-m\e_{0ij}\d(x-y)~.
\label{symp.mod}
\ee

The auxiliary field $\Phi$ is introduced in (\ref{bisd.lag}) for
re-expressing the Lagrangian in Eqn. (\ref{bisd1}) in a convenient form
and after embedding the above model, we will eliminate the $\Phi$ field.
Hence, in converting the second class constraints to the first class ones
using the BFT embedding we consider only the remaining constraints $\Pi_0
{~\rm and}~\W $.

Following the BFT procedure we enlarge the phase space by introducing a
pair of canonically conjugate variables $\psi$ and $\Pi_\psi$. Now we
modify the constraints $\Pi_0~ {\rm ~and~} \W$ such that they have
vanishing Poisson bracket among themselves.

The modified constraints read
\bea
\W_0 &=& \Pi_0 +\psi~,\label{mod.con1}\\
\W^\prime &=& \fr{2 \Phi}{\b^2} (f_0 + \Pi_\psi) -\fr{1}{m} \e_{0ij}\p^i
f^j \we 0~,\label{mod.con2}
\eea
and it is easy to see that the Poisson bracket between these two
constraints vanishes strongly. The general procedure of BFT requires the
construction of the embedded Hamiltonian which is in involution with
(\ref{mod.con1}) and (\ref{mod.con2}). Due to the non-linearity inherent
in this theory, the embedded Hamiltonian is an infinite series,
\be
H_{emb}=H_c -\Pi_\psi \W -\fr{\Phi \Pi_{\psi}^2}{\b^2}
+\fr{\psi}{2\Phi}\p_i (f^i\Phi)+.....~.
\ee 
In the improved BFT scheme, which make use of the fact that any function
of involutive combination of fields by itself is involutive, the embedded
Hamiltonian is constructed by replacing $f_0~{\rm and~} f_i$ by
corresponding gauge invariant combinations ${\bar f}_0~{\rm and~} {\bar
f}_i$ respectively.

Now corresponding to the original phase space variables $f_0~{\rm
and}~f_i$ we construct ${\bar f}_0~{\rm and}~{\bar f}_i$ which are in
strong involution with the modified constraints $\W_0~ {\rm
and}~\W^\prime$. Thus we get
\bea
{\bar f}_0 &=& f_0 +\Pi_\psi~,\label{gif0}\\ {\bar f}_i &=& f_i
+\fr{\b^2}{2\Phi} [ \p_i \psi -\psi \p_i (ln \Phi)]\label{gifi}~.
\eea
Next the embedded Hamiltonian has to be constructed. This is easily done
by the improved BFT scheme by replacing the original fields by involutive
combinations (\ref{gif0}, \ref{gifi}). Thus using (\ref{bisd.ham}) we get
the embedded Hamiltonian as
\be
{\cal H}_{emb} = -\fr{\Phi}{\b^2} {\bar f}_i{\bar f}^i -\Phi
-\fr{\b^4}{4\Phi} +\fr{\Phi}{\b^2} {\bar f}_0{\bar f}^0 - {\bar f}_0 {\bar
\W}~,
\label{bisd.emb}
\ee 
where ${\bar \W} = (2\fr{\Phi}{\b^2}{\bar f}^0 -\fr{1}{m}\e_{0ij}\p^i
{\bar f}^j)~.$ It is easy to see that the modified constraints $\W_0~{\rm
and}~\W_{i}^\prime$ (\ref{mod.con1}, \ref{mod.con2}) have vanishing
Poisson brackets with the embedded Hamiltonian (\ref{bisd.emb}). Note that
this embedded Hamiltonian reduces to (\ref{bisd.ham}), in the unitary
gauge, where the newly introduced fields $\a~{\rm and~}\Pi_\a$ are set to
zero.

The two equations of motion following from the embedded Hamiltonian
(\ref{bisd.emb}) are
\bea
\fr{2\Phi}{\b^2} {\bar f}_0 -\fr{1}{m}\e_{0ij}\p^i {\bar f}^j&=&0~,\\
\fr{2\Phi}{\b^2}{\bar f}_i-\fr{1}{m}\e_{i\m\n}\p^\m {\bar f}^\n&=&0~,
\eea
which can be expressed as
\be
\fr{2\Phi}{\b^2}{\bar f}_\m-\fr{1}{m}\e_{\m\n\l}\p^\n {\bar f}^\l=0~,
\label{eqn.ham}
\ee
in a covariant way. This equation imply $\p^\m (\Phi {\bar f}_\m)=0$.  A
gauge invariant solution of Eqn. (\ref{eqn.ham}) satisfying the condition
$\p^\m (\Phi {\bar f}_\m)=0$ is
\be
{\bar f}_\m =\fr{\b^2}{2\Phi}\e_{\m\n\l} (\p^\n A^\l -\p^\l A^\n)~.
\label{sol.ham}
\ee

On substituting this solution for ${\bar f}_\m$ in Eqn.~(\ref{bisd.emb}),
embedded Hamiltonian becomes,

\be
{\cal H}_{emb} = \fr{\b^2}{2 m^2 \Phi} F_{ij}F^{ij} -\fr{\b^2}{m^2 \Phi}
F_{0i}F^{0i} -\Phi -\fr{\b^4}{4\phi} -f_0{\tilde \W}~,\label{giham}
\ee
where ${\tilde \W} = (\e_{0ij}F^{ij} - \fr{\b^2}{m}\p^i(\fr{
F_{0i}}{\Phi})).$ Substituting for ${\bar f}_\m$ from Eqn.~(\ref{sol.ham})
in Eqn.~(\ref{eqn.ham}) we get the corresponding equation following from
the BICS theory.  With the identification $\fr{1}{m}F_{0i}=-E_i,{~\rm
and~}\fr{1}{{\sqrt 2}{m}}\e_{0ij}F^{ij}=B$, it is easy to see that the
above Hamiltonian (\ref{giham}) and constraints ${\tilde \W}~{\rm and~}
{\tilde \L}$ (${\tilde \L}$ is the constraint (\ref{sec2}) expressed in
terms of the solution for ${\bar f}_\m$ ) are the ones which follow from
the Lagrangian
\be
{\CL}=\Phi -\fr{\b^4}{4\Phi} - \fr{\b^2}{2 m^2\Phi} F_{\m\n}F^{\m\n}
+\fr{1}{2m}\e_{\m\n\l}A^\m\p^\n A^\l~.
\label{bimsc1.lag}
\ee
 Eliminating $\Phi$ from the above Lagrangian and redefining
$\fr{A_\m}{m}\to A_\m$ gives the BICS Lagrangian (\ref{bimcs1}).  This
shows the equivalence of the embedded version of the BISD model and the
BICS theory.

\section{ Duality Equivalence}\label{duality}

In this section we show that the BICS theory is related to BISD model
through duality. The duality equivalence between these two models is
obtained here in a Lagrangian formulation.

We derive this duality equivalence using Buscher's procedure 
\cite{{Buscher:1987pl},{Rocek:1992np}} of
constructing dual theories. Basically, this procedure consists of gauging
a global symmetry in the theory with a suitable gauge potential. To make
it equivalent to the original theory, we constrain the dual field strength
of the gauge potential to vanish by means of a Lagrange multiplier.
Integrating the multiplier field and  the gauge field, original action is
recovered. Instead, if one integrates the original field  and gauge
potential, the dual theory is obtained where the multiplier field becomes
the dynamical field.  This procedure has been used recently to show the
equivalence between a topologically massive gauge theory and different
St\"uckelberg formulations in $3+1$ dimensions\cite{us1}. The duality
relation obtained here between the BICS theory and BISD model nicely
complements the equivalence obtained between these theories in the
canonical formulation. We also use it to obtain the mapping between the
correlators of these two theories.

We start with the partition function for the BICS theory
\be
Z=\int DA_\m~ exp\left({~i\int d^3x {\CL}}\right)~,
\label{bimcsz}
\ee
where the Lagrangian
\be
{\CL}={\b}^2 \sqrt{1-\fr{1}{2{\b}^2}F_{\m\n}F^{\m\n}} +
\fr{m}{2}\e_{\m\n\l}A^\m\p^\n A^\l + \fr{1}{2}\e_{\m\n\l}F^{\m\n}J^\l~,
\label{lbics}
\ee
with $J^\l$ being the source term (which is coupled to dual field strength).
As in the last section,
we first linearize the Born-Infeld part of the above Lagrangian by
introducing an auxiliary field $\Phi$ .

Apart from the local $U(1)$ invariance (\ref{au1}), the action
corresponding to this Lagrangian is invariant under the global shift
symmetry of the vector field
\be
\d A_\m= \e_\m~,
\label{shift}
\ee
where $\e_\m$ is a constant.  As discussed earlier, the necessary
ingredient for dualization in Buscher's procedure is the presence of a
global symmetry. In order to gauge the global shift symmetry (\ref{shift})
of $A_\m$ in Chern-Simons term, it is convenient to linearize it using an
auxiliary vector field $P_\m$. Thus we re-express the Lagrangian 
(\ref{lbics}) as
\be
{\CL}=[1-\fr{1}{2{\b}^2}F_{\m\n}F^{\m\n}] \Phi +  \fr{\b^4}{4\Phi}
+\fr{m}{4}\e_{\m\n\l}P^\m F^{\n\l} -\fr{m}{8}\e_{\m\n\l}P^\m\p^\n P^\l +
\fr{1}{2}\e_{\m\n\l}F^{\m\n}J^\l~.
\label{linbimcs}
\ee

We convert the global symmetry (\ref{shift}) of the Born-Infeld
Chern-Simons Lagrangian into a local one by introducing a 2-form gauge
potential $~G_{\m\n}$. The source is coupled to (dual) field strength
rather than to field itself so as to keep the invariance under the
transformation (\ref{shift}). Following Buscher's procedure, we
constrain the dual field strength of the gauge field to be flat. This is
achieved by a multiplier field $\T$. Thus we get the Lagrangian
\bea
{\CL}&=& [1-\fr{1}{2{\b}^2}(F_{\m\n}-G_{\m\n})(F^{\m\n}-G^{\m\n})] \Phi
+\fr{\b^4}{4\Phi} +\fr{m}{4}\e_{\m\n\l}P^\m (F^{\n\l}-G^{\n\l}) \no\\ &+&
\fr{1}{2}\e_{\m\n\l}(F^{\m\n}-G^{\m\n})J^\l 
-\fr{m}{8}\e_{\m\n\l}P^\m\p^\n P^\l +\fr{1}{2}\e_{\m\n\l}\T \p^\m
G^{\n\l}~.
\label{busbimcs}
\eea
Here, under the local shift of $A_\m$, the gauge potential also transform
as $\d G_{\m\n}=(\p_\m\e_\n -\p_\n \e_\m)$ so that $(F_{\m\n}-G_{\m\n})$
( and hence the Lagrangian (\ref{busbimcs})~) is gauge invariant. Note
that under this transformation $P_\m$ and $\T$ remain invariant.

It is interesting to note that the action corresponding to the Lagrangian
(\ref{busbimcs}) is also invariant under $\d P_\m=\p_\m\l$ provided the
multiplier field  $\T$ undergoes a transformation $\d \T =-m\l/2$ while $\d
A_\m =0$ and $\d G_{\m\n}=0$. Thus we see that $\T$ undergoes a
compensating St\"uckelberg transformation.  Integrating $G_{\m\n}~{\rm
and}~A_\m$ from the partition function leads to the following Lagrangian
\bea
{\CL}&=&\Phi +\fr{\b^4}{4\Phi} +\fr{{\b}^2}{4\Phi} [ (\fr{m}{2}P_\m +\p_\m
\T) (\fr{m}{2}P^\m +\p^\m \T)] \no\\ &+& \fr{{\b}^2}{4\Phi} J_\m [J^\m + m
P^\m +2\p^\m \T] -\fr{m}{8}\e_{\m\n\l}P^\m\p^\n P^\l~.
\eea

By integrating out $\Phi$ from the above Lagrangian the partition function
reduces to
\be
Z=\int DP_\m D\T~ exp\left({~i \int d^3 x {\CL}}\right),
\label{bisdz}
\ee
where
\be
{\CL}= \b^2 \sqrt{ 1 +\fr{1}{\b^2}(\fr{m}{2}P^\m +\p^\m \T)^2
+\fr{1}{\b^2} J_\m (J^\m + m P^\m + 2\p^\m \T)}
-\fr{m}{8}\e_{\m\n\l}P^\m\p^\n P^\l~,
\label{lbisd}
\ee
showing the equivalence of the BICS theory to the St\"uckelberg
formulation of the BISD model. The above Lagrangian (\ref{lbisd}) is invariant 
under 
\bea
\d P_\m &=&\p_\m \l,\no\\
\d \T &=&-\fr{m}{2}\l~.
\label{lbisdst}
\eea
The field $\T$ introduced as a multiplier field in Eqn.~(\ref{busbimcs}) appears as the
St\"uckelberg field with correct compensating transformation in
Eqn.~(\ref{bisdz}). Note that 
in the last term in (\ref{lbisd}), the contribution from the
St\"uckelberg field vanishes.
Thus the model described by the Lagrangian (\ref{lbisd}) is invariant under 
the $U(1)$ transformation of $P_\m$ when the scalar field also undergoes a 
compensating transformation. Thus we see here that two theories which are 
related by duality have the same symmetry. The same feature was observed in the 
case of the 
duality equivalence of topologically massive $B{\wedge}F$ theory in $3+1$ dimensions 
to two different massive gauge theories \cite{us1}.
In the absence of the sources terms
$(i.e., J_\m=0)$, by setting the gauge condition $\T=0$ and with the
identification $\fr{m P_\m}{2}=f_\m$, the above Lagrangian reduces to that
BISD model (\ref{bisd1}). Thus using Buscher's procedure we have shown
that the BICS theory is equivalent to the gauge invariant form of the 
BISD model.
We see here that the BISD model described by (\ref{bisd1}) is the gauge fixed 
version of  the dual model obtained by Buscher's procedure from BICS theory.

>From the partition functions of Eqn. (\ref{bimcsz} and \ref{bisdz}), by
functional differentiation with respect to corresponding source terms and
then setting sources to be zero, we get the following mapping between the
correlators

\bea
\left <~F_{\m}^{*}~ F_{\n}^{*}~\right >_{BICS}~&\rla&~  
\fr{1}{4} ~\left <~(m P_\m +2\p_\m \T) ~(m P_\n +2\p_\n \T)~\right
>_{BISD} \no\\ &-& i g_{\m\n}\d(x-y)~,
\label{cmap}
\eea
where $F_{\m}^{*}$ is the dual of $F^{\n\l}$ and the last term in the
above is a non-propagating contact term. It is interesting to note that
the (dual) field strength of the BICS theory is equivalent to a gauge
invariant combination of fields in the gauge invariant form of BISD model
( This result is also true in the case of the MCS theory and the SD
model). In particular, remarkably there is no $\b$ dependence in the
operator equivalence.

\section{Bosonization}\label{bose}

As discussed in the introduction, the MCS gauge theory has been shown to be
equivalent to the massive Thirring model. The operator correspondence at
the level of correlators  between the fermionic current and the dual gauge
field strength of MCS has been shown 
\cite{{Redlich:1984prd},{Shaji:1990ij},{Shankar:1991ij}}. 
Also the commutators between the
components of the fermionic current has been evaluated using this
correspondence and the Dirac brackets of MCS theory. Further, non-zero
Schwinger term is found to result in the $[j_0 (x), j_i(y)]$ commutators,
where $j_\m$ is the fermionic current. It is then clearly of considerable
interest to extend the previous study to the BICS case and investigate the role
of non-linearity in the bosonization of the Born-Infeld theories.

In this section, we first start from the the  BIT model (\ref{bith1}) and
show that to leading order in the fermion mass $\mu$, 
its partition function is equivalent to that of BICS theory. 
This is also
re-derived starting with an interpolating master Lagrangian. We also
provide the bosonization rules relating the fermion current correlators of
BIT to (dual) field strength correlators of the BICS theory. 
We find that this mapping is
$\b$ independent. We also use this correspondence to obtain the current
algebra for the fermionic theory. Since this requires the Dirac brackets
for the BICS theory which to the best of our knowledge have not been
evaluated so far, hence we have calculated them in the Appendix.  Using
these Dirac brackets, the current algebra relations are evaluated and are
found to be $\b$ dependent. As expected, in the limit $\b\to\infty$, the
MCS-Thirring model results are reproduced.

\subsection{ Non-Polynomial Thirring model}\label{npt} 

We start with the partition function of a massive Dirac fermion with a
non-polynomial Thirring type current-current coupling whose partition
function is
\be
Z_{T}=\int D{\bar \Psi}D\Psi exp~\left({i~\int d^3 x {\CL}}\right)~,
\label{fpart}
\ee
where ${\CL}$ is given by Eqn. (\ref{bith1}). 

The non-polynomial interaction term in this
Lagrangian is linearized by introducing an auxiliary field $\chi$ to
express (\ref{bith1}) as
\be
L= {\bar \Psi}(i\c_\m \partial^\m-\m)\Psi +\chi +\fr{\b^4}{4\chi}
-\fr{\chi\l}{\b^2} j_\m j^\m~.
\label{linbith}
\ee
Next we linearize the quadratic term $j_\mu j^\m $ by an auxiliary vector
field $ f_\m$ and write the Lagrangian as
\be
{\CL}={\bar \Psi}(i\c_\m \partial^\m-\m)\Psi +\chi +\fr{\b^4}{4\chi} +
\fr{\l \chi}{\b^2} (\fr{1}{4}f_\m f^\m - f_\m j^\m)~.
\label{folag}
\ee
Using this Lagrangian in (\ref{fpart}), we integrate out the Dirac field
and use the well-known result for the 
evaluation of  the determinant of the Dirac
operator  in a gauge invariant regularization 
\cite{Redlich:1984prd}:
\be
ln~ det~(i\g_\m \p^\m -\m -g f_\m \g^\m ) = \pm
\fr{i g^2}{8\pi}\int d^3 x \e_{\m\n\l}f^\m \p^\n f^\l +O(\fr{1}{\m})+...~.
\label{detfem}
\ee
The leading order term is the odd parity Chern-Simons term and 
as is well known, in the limit 
$\m\to\infty$, only this term survives.
This applied to the present case gives
\bea
ln~ det~(i\g_\m \p^\m -\m -\fr{\chi\l}{\b^2}f_\m \g^\m ) &=& \pm
\fr{i\l^2}{8\b^4\pi}
\int d^3 x \chi \e_{\m\n\l}f^\m \p^\n (\chi f^\l) + O(\fr{1}{\m})+...\no\\
&=& \pm\fr{i\l^2}{8\b^4\pi}
\int d^3 x \chi^2 \e_{\m\n\l}f^\m \p^\n f^\l + O(\fr{1}{\m})+...~.
\eea
As a result, in the limit $\m\to\infty$, the partition function becomes
\be
Z=\int D\chi Df_\m exp~i\left(~\int d^3 x {\CL}_{eff}\right)~,
\label{effz1}
\ee
where
\be
{\CL}_{eff}= \chi +\fr{\b^4}{4\chi} + \fr{\b^2}{4\chi} g_{\m}g^\m
\pm \fr{\l}{8\pi} \e^{\m\n\l}g_{\m}\p_\n g_{\l}~, 
\label{efflag}
\ee
(we have redefined $\fr{\sqrt{\l}\chi}{\b^2} f_\m = g_{\m} $). 
We identify the above Lagrangian (\ref{efflag}) with that given in
(\ref{bisd.lag}) describing BISD model (in the same spirit as 
\cite{{Redlich:1984prd},{Fradkin:1994pl},{Rabin:1995pl}}).
Now we integrate out the auxiliary field $\chi$ and resulting Lagrangian is
\be
{\CL}_{BISD} =\b^2\sqrt {1+\fr{1}{\b^2}g_{\m} g^{\m}} \pm
\fr{\l}{8\pi} e^{\m\n\l}g_{\m} \p_\n g_{\l}~, 
\label{fbbisd}
\ee
which is the BISD action given in (\ref{bisd1})
with the identifications $m=\mp\fr{4\pi}{\l}$. Thus we have shown that  
to order
$\fr{1}{\m}$, the Thirring model partition function is equivalent to 
that of BISD model.
Now using the 
equivalence discussed in the previous section, we can identify this 
Lagrangian, with the BICS Lagrangian (\ref{bimcs1}) and thus we conclude that
\be
Z_{T}\equiv Z_{BICS}~.
\ee

\subsection{ Master Lagrangian}

Next we derive the same result from a master Lagrangian, as has been done
in the usual MCS to Thirring model equivalence.

The master Lagrangian we start with is
\bea
L ={\bar \Psi}(i\c_\m \partial^\m-\m)\Psi -\a f_\m j^\m
-\b^2\sqrt{1-\fr{1}{2\b^2} F_{\m\n}F^{\m\n}}\no\\
+\fr{1}{4}\e_{\m\n\l}f^\m F^{\m\l}+\fr{1}{4}\e_{\m\n\l}F^{\m\n} J^\l~.
\label{masterlag}
\eea
In the above $J_\m$ is the source for the dual of the field strength
$F_{\m\n}=(\p_\m A_\n-\p_\n A_\m),$ $\a$ is a dimensionful constant,
$j_\m$ is the conserved fermionic current, ${\bar\Psi}\c_\m\Psi$. The
Lagrangian (\ref{masterlag}) is invariant under (i) the local $U(1)$
transformation of gauge potential $A_\m$ (ii) under another independent
local $U(1)$ gauge transformation where $\Psi$ and $f_\m$ are minimally coupled
and (iii) global shift symmetry of $A_\m$.

Next we implement the Buscher's procedure to re-express this theory into
an equivalent one. Note that in contrast to section III, here we have a
mixed Chern-Simons term involving 1-forms $f$ and $A$. To apply the
Buscher's procedure we make use of the shift symmetry 
($\d A_\m =\e_\m$, where $\e_\m$ is constant) 
of the Lagrangian (\ref{masterlag}).
Note that in (\ref{masterlag})
it was necessary to
couple the source to the dual field strength of $A_\m$ so as to have  
the shift symmetry.

As in the previous section, we first linearize the Born-Infeld term using
an auxiliary $\Phi$ field.  The Lagrangian which is invariant under the
local shift of $A_\m$ is given by
\bea
L&=&{\bar \Psi}(i\c_\m \partial^\m-\m)\Psi -\a f_\m j^\m +\Phi +\fr{\b^4}{4\Phi}
-\fr{\Phi}{2\b^2} (F_{\m\n}-G_{\m\n})(F^{\m\n}-G^{\m\n}) \no\\&+&\fr{1}{4} 
\e_{\m\n\l}(F^{\m\n}-G^{\m\n})(f^\l~+~J^\l)
-\fr{1}{4}\e_{\m\n\l}G^{\m\n}\p^\l \T~,
\label{bfbus}
\eea where $G_{\m\n}$ is a 2-form gauge potential and $\T$ is the
multiplier field.  Note that as in the previous section, we have
constrained the dual field strength of $G_{\m\n}$ to be flat.  We
integrate out $G_{\m\n}$ and $A_\m$ fields from the partition
function corresponding to (\ref{bfbus}). Thus we get
\be
Z=\int D{\bar \Psi}D\Psi Df_\m D\T D\Phi~exp~\left({i\int d^3 x
~L_{eff}}\right)~,
\label{effz}
\ee
where
\bea
L_{eff}&=& {\bar \Psi}(i\c_\m \partial^\m-\m)\Psi -\a f_\m j^\m +\Phi
+\fr{\b^4}{4\Phi} +\fr{\b^2}{16\Phi} (f_\m +\p_\m \T)(f^\m+\p^\m
\T)\no\\&+&\fr{\b^2}{8\Phi} (f_\m +\p_\m\T)J^\m +\fr{\b^2}{16\Phi}J_\m J^\m~.
\eea
Next, the integrations over $f_\m,~{\rm and}~\T$ which are Gaussian are carried
out with the gauge condition $\T=0$ \footnote{ We can as well fix the
gauge condition to be $\p_\m f^\m=0$ and then integrate over $f_\m ~{\rm
and}~\T.$} and then we integrate out $\Phi$ also to get the partition
function
\be
Z[J_\m] = \int D{\bar \Psi}D\Psi ~exp~\left({i\int d^3x ~{\CL}}\right)~,
\label{mtmz}
\ee
where
\be
{\CL}=  {\bar \Psi}(i\c_\m \partial^\m-\m)\Psi -\a J_\m j^\m +
\b^2\sqrt{1-\fr{\a^2}{\b^2}j_\m j^\m} ~.
\label{lag.mtm}
\ee
 In the absence of source $J_\m$, the Lagrangian (\ref{lag.mtm})
reduces to that of the BIT model (\ref{bith1}) with the identification
$\a^2=\l$.

Instead of integrating the bosonic variables from the partition function
of the master Lagrangian (\ref{masterlag}), one can integrate ${\bar
\Psi}~{\rm and}~\Psi$. Using Eqn.~(\ref{detfem}), we get 
\be
Z[J_\m]=\int DA_\m Df_\m exp~\left({i\int d^3 x {\CL}}\right)~,
\label{bose.z}
\ee
where
\bea
{\CL}&=&\b^2 \sqrt { 1-\fr{1}{2\b^2} F_{\m\n}F^{\m\n}}
+\fr{1}{4}\e_{\m\n\l}f^\m F^{\n\l} + +\fr{1}{2}\e_{\m\n\l}F^{\m\n}
J^\l\no\\ & -&\fr{\a^2}{8\pi}\e_{\m\n\l}f^\m\p^\n f^\l + O(\fr{1}{\m})+...~.
\label{bose.lag}
\eea
Since the partition functions in Eqns. (\ref{mtmz}) and (\ref{bose.z}) are
obtained from the same master Lagrangian, this shows their equivalence.

Integrating out $f_\m$ from (\ref{bose.z}) with the gauge condition $\p_\m
f^\m=0$, we get the partition function to
be
\be
Z=\int DA_\m exp~i~\left(\int d^3x {\CL}\right)~,
\label{part.bics}
\ee
where
\be
{\CL}=\b^2 \sqrt { 1-\fr{1}{2\b^2} F_{\m\n}F^{\m\n}}
+\fr{m}{2}\e_{\m\n\l}A^\m \p^\n A^\l +\fr{1}{2}\e_{\m\n\l}F^{\m\n}J^\l~,
\label{lagbicsbus}
\ee
with the identification  $\fr{4\pi}{\a^2} =m$~. For $J_\m=0$, this gives
\be
Z_{T}\equiv Z_{BICS}~.
\ee

Using the fact that (\ref{mtmz}) and (\ref{part.bics}) are equivalent, one
can easily derive the mapping between the correlators of these two models
by taking the functional derivatives with respect to the source $J_\m$.
Thus we get
\be
\left <j_{\m_1}(x_1)..~j_{\m_n} (x_n) \right>_{BIT}= \fr{1}{\a^{2n}} \left <
F_{\m_1}^*(x_1)..~~F_{\m_n}^*(x_n)\right>_{BICS}~,\label{fbmap1}
\ee
where $F_{\m}^*$ is the dual field strength corresponding to $A_\m$.
This implies the following operator correspondence between the fermionic
current and the dual field strength:
\be
j_\m = \fr{1}{2\a}\e_{\m\n\l}F^{\n\l}~.
\label{dffc}
\ee
Here we note that the mapping (\ref{fbmap1}) is independent of the $\b$
parameter and thus same as that obtained in the case of MCS - massive
Thirring model.

\subsection{Current Algebra}

In this subsection we extend the analysis of bosonization by studying the
algebra of fermionic currents by using the operator equivalence
(\ref{dffc}).

We see that with the identification (\ref{dffc}), the term
$\sqrt{1-\fr{1}{2\b^2}F_{\m\n}F^{\m\n}}$ in BICS goes to
$\sqrt{1-\fr{\a^2}{2\b^2}j_\m j^\m}$ and if 
further $\a^2=\l$ then this is precisely the
non-linear current-current term in Eqn.~(\ref{bith1}).  From the relation
(\ref{dffc}) between the fermionic current and the dual field strength we
can obtain the potential $A_i$ in terms of the components of the current
$j_\m$. Then using the Dirac brackets evaluated in the Appendix A
we get the various commutators between the currents. From
Eqn. (\ref{dffc}) we have for the components
\bea 
j_0&=& \fr{1}{\a}\e_{0ij}\p^i A^j~,\label{com1}\\
j_i&=&-\fr{1}{2\a}\e_{0ij}F^{0j}~.
\label{com2}
\eea
Inverting these relations we get
\bea
A_i&=&\fr{\a}{\nabla^2}\e_{0ij}\p^j j^0~,\label{aij}\\
\Pi_i&=&-\fr{\a}{R}\e_{0ij}j^j -\fr{m\a}{2\nabla^2}\p_i j^0\label{pii}~,
\eea
where $\Pi_i$ is the conjugate momentum corresponding to $A_i$ and $R(x)
=\sqrt {1 -\fr{\a^2}{2\b^2}j_\m j^\m}.$ In obtaining (\ref{aij}) we have
used the condition $\p_i A^i=0$. Here we see that the conjugate
momentum $\Pi_i$ depends on $R$ and hence on $\b^2$. For the second
Eqn.~(\ref{pii}) we have used the defining relation of the conjugate
momenta $\Pi_i$  (viz: $\Pi_i =-\fr{F_{0i}}{R} +\fr{m}{2}\e_{0ij}A^j$).

Now using the non-vanishing Dirac brackets of BICS theory (see appendix,
Eqns.~\ref{adb1},~\ref{adb2}), we evaluate the commutators between
the different components of the fermionic current. Thus we get 
\be
\left [~j_0(x),~j_0(y)~\right ] = \left [ \fr{1}{\a}\e_{0ij}\p^i A^j(x),~
\fr{1}{\a}\e_{0lm}\p^l A^m (y)\right ] = 0~.\label{comm1}
\ee
Using the above commutator (\ref{comm1}) we get
\be
\{A_i(x), \Pi_j(y)\}^*= \{\fr{\a}{\nabla^2}\e_{0ij}\p_{(x)}^j
j^0{(x)},~-\fr{\a}{R(x)}\e_{0ij}j^j{(y)} \}^*~.
\label{eq1}
\ee
Using the Dirac bracket (\ref{adb1}) and with some algebra, from the above
Eqn.~(\ref{eq1}) we get
\be
\left [~j_0(x),~\fr{1}{R(y)} j_i(y)~\right ] = i \fr{1}{\a^2}
\p_{i}^{(x)}\d(x-y)~. 
\label{comm2}
\ee
Similarly from
\be
\{\Pi_i(x), \Pi_j(y)\}^*= \{-\fr{\a}{R(x)}\e_{0il}j^l{(x)}
-\fr{m\a}{2\nabla^2}\p_{i}^{(x)} j^0{(x)} ,~ -\fr{\a}{R(x)}\e_{0jm}j^m (y)
-\fr{m\a}{2\nabla^2}\p_{j}^{(y)} j^0{(y)} \}^*~,
\ee
using the above commutators (\ref{comm1}, \ref{comm2}), and the Dirac
bracket (\ref{adb2}) we get
\be
\left [~\fr{1}{R(x)}j_{i}{(x)},~\fr{1}{R(y)}j_{j}{(y)}~\right ]
=-i\fr{m}{\a^2}\e_{oij}\d(x-y)~.
\label{comm3}
\ee
Here we see that the commutators (\ref{comm2}, \ref{comm3}) 
are $\b$ dependent (note $R$ is $\b$-dependent)
even though the correlator mapping (\ref{fbmap1}) is
$\b$-independent. In the limit $\b^2\to\infty$,  from (\ref{comm2},
\ref{comm3}) we get 
\bea
\left [~j_{0}{(x)},~j_{i}{(y)}~\right ] &=& i \fr{1}{\a^2} \p_{i}^{(x)}\d(x-y)
-\fr{\a^2}{4\b^2} \left [ j_{0}{(x)}, j_\m j^\m j_{i}{(y)}\right ] +
O(\fr{1}{\b^4})~+..~, \label{com2a}\\
\left [j_{i}{(x)},~j_{j}{(y)}~\right ]
&=&-i\fr{m}{\a^2}\e_{oij}\d(x-y)\no\\& -&
\fr{\a^2}{4m} \left [ j_{i}{(x)}j_\m
j^\m,~j_{j}{(y)}\right ] -\fr{\a^2}{4\b^2}\left [j_{i}{(x)}, j_\m j^\m
j_{j}{(y)}\right]~+~ O(\fr{1}{\b^4})~+..~.
\label{com3a}
\eea

Note that the operator correspondence is $\beta$ independent, 
since it is a general feature of bosonization. However, the current algebra, which is 
an observable of the theory, depends on the details of the theory and hence on $\beta$.

It is easy to see from the above Eqns.~(\ref{com2a}, \ref{com3a}) that 
as $\b^2 \rightarrow \infty$ , the above commutators
(\ref{comm1},\ref{comm2},\ref{comm3}) reduce to the corresponding
ones of the massive Thirring model evaluated using the Dirac
brackets of MCS theory\cite{Rabin:1995pl}.

In order to understand the free theory limit, it is convenient to scale the fields 
$A_\m~\to~\a A_\m$ so that (\ref{dffc}) is $\a$ independent. Then as 
$\a~\to~0$,
only Chern-Simons term survives in the action while 
the Fermionic sector becomes free theory
and gives the same prescription of the current bosonization.

\section{Conclusion}\label{con}

In this paper we have made a comprehensive study of the BICS theory and
the SDBI model and shown their equivalence in a variety of ways.  We
have extended the earlier  studies on the MCS theory and  the SD model to the
Born-Infeld case. One of the motivation is to investigate the application
of the known techniques to non-linear theories.

We started with the SDBI theory, which has only second-class
constraints, and following the improved BFT scheme we obtained the gauge
invariant Hamiltonian and also converted the constraints to first class.
The resulting 
theory was shown to be equivalent to the BICS theory. 
This demonstrates that the improved BFT procedure is also applicable to 
highly non-linear theories. Previously
this method has been successfully applied to other non-linear
theories like massive Yang-Mills theory\cite{Narayan:1995an}. Apart from
its intrinsic interest this study also provides yet another example of the
application of this scheme to a different type of nonlinear theory.

Next we showed that the equivalence between these theories, viz; BICS and
BISD , is actually a duality equivalence thereby extending the approach of
Buscher's procedure to relate two {\it non-linear dual theories} . 

We have also provided the operator
correspondence between the fermionic current of the BIT and the dual gauge
field of the BICS at the level of correlators. This has been used to
calculate the commutators between different components of the fermionic
current.
For this purpose, we
have calculated the Dirac brackets of the BICS theory using symplectic
quantization scheme in Appendix A and this by itself is an interesting new
result.  The current algebra is found to be $\b$ dependent and the leading
order correction term in $\fr{1}{\b^2}$ has been computed. 
As expected, in the limit
$\b^2\to\infty$, the results corresponding to massive Thirring model are
recovered. It is interesting to note that the coupling constant 
$\a^2$ that appears linearly in BIT, appears inversely as Chern-Simons 
mass of BICS. This can be of use in studing the non-perturbative 
aspects of BICS. 

\noindent {\bf Acknowledgments}

\noindent One of us (EH) would like to thank Institute of
Physics, Bhubaneswar for the warm hospitality extended to him during his
visit to the institute where a part of this work was done.

\newpage
\appendix
\section {Symplectic Quantization}\label{a1}
In this appendix we compute the Dirac brackets of the BICS theory using
the symplectic quantization scheme.

The symplectic quantization scheme of Faddeev and Jackiw
\cite{{Faddeev:1988pr},{Wotzasek:1992ij},{Neto:1992ij}}
has recently been studied and applied to different models. Primary
constraints of the Dirac procedure do not appear in this Lagrangian
scheme. Here one starts with the Lagrangian which is first
order in time derivatives. One then has to invert the symplectic matrix to
obtain the Dirac brackets. If the system has true constraints, then the
symplectic matrix become singular. In this case the configuration space is
enlarged by introducing multiplier fields corresponding to each of the
constraints and the constraints are introduced back in to the Lagrangian
using them. After incorporating all the constraints, the symplectic matrix
can still be singular signaling the gauge invariance of the theory. At
this stage the gauge fixing conditions are introduced so as to make the
symplectic matrix non singular and from its inverse Dirac brackets are
obtained.

Here we apply this scheme to the BICS theory described by the Lagrangian
(\ref{bimcs1}) and obtain the Dirac brackets. We start with a first order
form of (\ref{bimcs1}) as given by
\be
{\CL}=\Pi_i \p^0 A^i - R \Pi_i \Pi^i +\fr{m^2}{4} R A_i A^i +\b^2 R +
(\p^i \Pi_i +\fr{m}{4}\e_{0ij}F^{ij})A^0~.
\label{symlag}
\ee
Here as is common in the symplectic scheme, $\Pi_i$ is not identified with the
conjugate momentum corresponding to the field $A_i$ and
$R=\sqrt{1-\fr{1}{2\b^2}F_{\m\n}F^{\m\n}}$. The above Lagrangian
(\ref{symlag}) is of the form
\be
L=a_i \p^0 \zeta^i - V(\zeta)~,
\label{1st.lag}
\ee
and in that case the symplectic matrix is defined as
\be
f_{ij}=\p_i a_j -\p_j a_i\label{2form}~,
\ee
where the derivatives are taken with respect to $\zeta$. In our case,
using (\ref{symlag}), we get the $a_i$ to be
\be
a_{0}^A=0~~~~~~a_{i}^A=\Pi_i,~~~a_{i}^\Pi=0~,
\ee
where $a_{0}^A$ stand for the coefficient of $\p^0 A_0$. Then the only
non-vanishing elements of the symplectic matrix are
\be
f_{ij}^{A\Pi}(x,y) = \fr{\p a_{j}^\Pi}{\p A_i} -\fr{\p a_{i}^A}{\p \Pi_j}
=-\d_{ij}\d(x-y) = - f_{ij}^{\Pi A} (x,y)~.
\ee
Thus we get the symplectic matrix to be
\bea
 f_{ij}(x,y)=\left (
\begin{array}{ccc}
0~&~0~&~0 \cr 0~&~0~&~-\d_{ij} \cr 0~&~\d_{ij}~&~0
\end{array}\right )\d(x-y)~,
\eea
which is singular, showing the constrained nature of the theory.  These
constraints are obtained by solving for the zero modes of $f_{ij}$
from\footnote{Since the Euler-Lagrange equation following from
(\ref{1st.lag}) is $f_{ij}\p^0\zeta^j =\fr{\p~V(\zeta)}{\p \zeta_i}$, the
zero mode of $V(\zeta)$ is related to that of $f_{ij},$ where $f_{ij}$ is
defined in (\ref{2form}).}
\be
\int dx dy \left [ a\fr{\d}{\d A_0} + b_i\fr{\d}{\d A_i} +c_i \fr{\d}{\d
\Pi_i}\right ] V =0~,
\ee
where $V = R \Pi_i \Pi^i -\fr{m^2}{4} R A_i A^i -\b^2 R - (\p^i \Pi_i
+\fr{m}{4}\e_{0ij}F^{ij})A^0~.$ From this we get the constraint
\be
\p_i \Pi^i +\fr{m}{2}\e_{0ij}F^{ij} =0~.
\label{gauss}
\ee
Now we introduce this constraint into the kinetic part of the Lagrangian
using a multiplier field $\eta$. In this way, the modified Lagrangian
takes the form
\be
{\CL}^1 = \Pi_i \p^0 A^i -\p^i \eta [ \p^0 \Pi_i +\fr{m}{2}\e_{0ij}\p^0
A^j] - R \Pi_i \Pi^i +\fr{m^2}{4} R A_i A^i +\b^2 R~,
\label{fjlag1}
\ee
where we have absorbed $A_0(\p_i \Pi^i +\fr{m}{2}\e_{0ij}F^{ij})$ into
 the second term of the above Lagrangian (\ref{fjlag1}). From
(\ref{fjlag1}) we get
\be
a_{i}^A = \Pi_i +\fr{m}{2}\e_{0ij} \p^j\eta,~~~a_{i}^\Pi =-\p_i\eta~,
\ee
and hence the symplectic matrix is
\bea 
f_{ij}(x,y) = \left (
\begin{array}{ccc} 
0~&~-\d_{ij}~&~-\fr{m}{2}\e_{0ij}\cr
\d_{ij}~&~0~&~\p_i \cr
\fr{m}{2}\e_{0ij}\p^j~&~-\p_i~&~0
\end{array} \right) \d(x-y)~,
\eea
which is also singular. But this does not give any new constraint showing
the gauge symmetry of the theory. So we introduce the gauge fixing
condition $\p_i A^i$ using a multiplier field $\l$ into the kinetic part
of the Lagrangian (\ref{fjlag1}) to get the modified Lagrangian
\be
{\CL}^{(2)} = (\Pi_i +\fr{m}{2}\e_{0ij}\p^j\eta -\p_i \l ) \p^0 A^i-\p_i\eta
\p^0 \Pi^i -R \Pi_i \Pi^i +\fr{m^2}{4} R A_i A^i +\b^2 R~.
\label{fj2}
\ee
The non-singular symplectic matrix following from the above Lagrangian
(\ref{fj2}) is
\bea f_{ij}^{(2)} (x,y)=\left (
\begin{array}{cccc}
0~&~-\d_{ij}&~-\fr{m}{2}\e_{0ij}\p^j~&~\p_i\cr
\d_{ij}~&~0~&~\p_i~&~0\cr
\fr{m}{2}\e_{0ij}\p^j ~&~-\p_j~&~0~&~0\cr
-\p_i~&~0~&~0~&~0
\end{array} \right) \d(x-y)~,
\eea
from the inverse of which we get the following non-vanishing Dirac brackets 
\bea
\left \{ A_i(x),~\Pi_j(y) \right \}^*&=&(\d_{ij}
+\fr{\p_i\p_j}{\nabla^2})\d(x-y),\label{adb1}\\
\left \{\Pi_i(x),~\Pi_j(y) \right \}^*&=& -\fr{m}{2\nabla^2}
(\e_{0mi}\p^m\p_j-\e_{0mj}\p^m\p_i)\d(x-y) .
\label{adb2}
\eea
In the above (\ref{adb1}, \ref{adb2}), all the derivatives are with
respect to $x$. Here we note that the  Dirac brackets are independent
of $\b^2$ and hence are the same as in the MCS theory. This is a
reflection of the fact that the gauge symmetry in the BICS theory 
and MCS theory is the same. 
The nonlinearity of  the BICS Lagrangian 
shows up only in the definition of $\Pi_i$ .

\end{document}